\documentclass[11pt]{article}
\pdfoutput=1

\usepackage{jheppub} 

\usepackage{verbatim} %for comments
\usepackage{float}
\restylefloat{table}
\usepackage[all]{xy}
\usepackage{pdflscape}
\usepackage{tikz}
\usepackage{array}
\usepackage{youngtab}
\usepackage{multirow}
\usepackage{color}
\usepackage{ulem}
\usepackage{cleveref}

\makeatletter

\newcommand{\be}{\begin{equation}}
\newcommand{\ee}{\end{equation}}
\newcommand{\ba}{\begin{aligned}}
\newcommand{\ea}{\end{aligned}}

\newcommand{\bea}{\begin{eqnarray}}
\newcommand{\eea}{\end{eqnarray}}
\newcommand{\tW}{{\rm w}}

\def\unit{{1\kern-.65ex {\rm l}}}
\def\1{{1\kern-.65ex {\rm l}}}

\def\slash#1{\ooalign{\hfil/\hfil\crcr$#1$}}

\makeatother

\begin{document}

% format
% \baselineskip=14pt  % a la harvmac
%\baselineskip=14pt
% \parskip 5pt plus 1pt 
\numberwithin{equation}{section}  % make eq labels (sec.num)
%\allowdisplaybreaks  % allow page breaks in displayed eqs

% print date, time and filename 
%\pagestyle{myheadings}
%\markright{{\tt \jobname.tex} -- \today{} \now}

%%%%%%%%%%%%%%%%%%%%%%%%%%%%%%%%%%%%%%%%%%%
%%%        TITLE BEGINS HERE
%%%%%%%%%%%%%%%%%%%%%%%%%%%%%%%%%%%%%%%%%%%

\vspace*{0.8cm} 
\begin{center}
 {\LARGE  The Gravitational Sector of 2d $(0,2)$ F-theory Vacua\\}

 \vspace*{1.8cm}
{Craig Lawrie$\,^1$, Sakura Sch\"afer-Nameki$\,^2$, and Timo Weigand$\,^1$}\\

 \vspace*{1.2cm} 
{\it $^1$ Institut f\"ur Theoretische Physik, Ruprecht-Karls-Universit\"at,\\
 Philosophenweg 19, 69120 Heidelberg, Germany }\\
 {\tt {gmail:$\,$craig.lawrie1729}} \\
 {\tt {t.weigand\phantom{@}thphys.uni-heidelberg.de}}\\
  
\bigskip
{\it $^2$ Mathematical Institute, University of Oxford \\
 Woodstock Road, Oxford, OX2 6GG, UK}\\
  {\tt {gmail:$\,$sakura.schafer.nameki}}\\
\vspace*{0.8cm}
\end{center}
\vspace*{.5cm}
%
% Abstract
\noindent
F-theory compactifications on Calabi--Yau fivefolds give rise to
two-dimensional $N=(0,2)$ supersymmetric field theories coupled to gravity. We
explore {the} dilaton supergravity defined by the moduli sector of such
compactifications.  The massless moduli spectrum is found by uplifting Type
IIB compactifications on Calabi--Yau fourfolds.  This spectrum matches
expectations from duality with M-theory on the same elliptic fibration. The
latter defines an $N=2$ Supersymmetric Quantum Mechanics related to the 2d
$(0,2)$ F-theory supergravity via circle reduction. Using our recent results
on the gravitational anomalies of duality twisted D3-branes wrapping curves in
Calabi--Yau fivefolds we show that the {F-theory spectrum} is anomaly free. We
match the classical Chern-Simons terms of the M-theory Super Quantum Mechanics
to one-loop contributions to the effective action by $S^1$ reduction of the
dual F-theory.

%%%%%%%%%%%%%%%%%%%%%%%%%%%%%%%%%%%%%%%%%%%
%%%           TITLE ENDS HERE
%%%%%%%%%%%%%%%%%%%%%%%%%%%%%%%%%%%%%%%%%%%

\pagenumbering{gobble}
\newpage
\pagenumbering{arabic}
%%%%%%%%%%%%%%%%%%%%%%%%%%%%%%%%%%%%%%%%%%%
%%%        MAIN TEXT BEGINS HERE
%%%%%%%%%%%%%%%%%%%%%%%%%%%%%%%%%%%%%%%%%%%

%%%%%%%%%%%%%%%%%%%%%%%%%%%%%%%%%%%%%%%%%%%%%%%%%%
%%%%%%%%%%%%%%%%%%%%%%%%%%%%%%%%%%%%%%%%%%%%%%%%%%
%%%%%%%%%%%%%%%%%%%%%%%%%%%%%%%%%%%%%%%%%%%%%%%%%%
\tableofcontents
%%%%%%%%%%%%%%%%%%%%%%%%%%%%%%%%%%%%%%%%%%%%%%%%%%

%%%%%%%%%%%%%%%%%%%%%%%%%%%%%%%%%%%%%%%%%%%%%%%%%%
%%%%%%%%%%%%%%%%%%%%%%%%%%%%%%%%%%%%%%%%%%%%%%%%%%
%%%%%%%%%%%%%%%%%%%%%%%%%%%%%%%%%%%%%%%%%%%%%%%%%%

%%%%%%%%%%%%%%%%%%%%%%%%%%%%%%%%%%%%%%%%%%%%%%%%%%
%%%%%%%%%%%%%%%%%%%%%%%%%%%%%%%%%%%%%%%%%%%%%%%%%%
\section{Introduction and Summary}
%%%%%%%%%%%%%%%%%%%%%%%%%%%%%%%%%%%%%%%%%%%%%%%%%%
%%%%%%%%%%%%%%%%%%%%%%%%%%%%%%%%%%%%%%%%%%%%%%%%%%

Chiral two-dimensional string compactifications  have recently attracted
revived interest
\cite{Franco:2015tna,Franco:2015tya,Schafer-Nameki:2016cfr,Franco:2016nwv,Apruzzi:2016iac,Franco:2016qxh,Apruzzi:2016nfr}.  
These works are in part motivated by the desire to better understand
the rich class of 2d $N=(0,2)$ supersymmetric conformal field theories (SCFTs)
and the hope to construct new examples of such systems as the infrared fixed
points of chiral two-dimensional string vacua. 
{With this motivation in mind, F-theory has
emerged as an ideal} framework for engineering and potentially even
classifying {SCFTs} in various dimensions, as
exemplified by the powerful results of \cite{Heckman:2013pva,Heckman:2015bfa}
on 6d $N=(1,0)$ {SCFTs}. 

The systematic investigation of the 2d effective theory obtained by
compactifications of F-theory on Calabi--Yau fivefolds has been initiated in
\cite{Schafer-Nameki:2016cfr,Apruzzi:2016iac}. Such theories comprise three
different chiral subsectors.  The particular focus in
\cite{Schafer-Nameki:2016cfr,Apruzzi:2016iac} was on the gauge dynamics
resulting from the partially topologically twisted theory on 7-branes wrapping
complex K\"ahler {six}-cycles on the base of the elliptic fibration.
Besides the 7-brane sector, F-theory compactifications on Calabi--Yau
fivefolds necessarily include{, due to tadpole constraints,} D3-branes wrapping holomorphic curves on
the base of the fibration \cite{Schafer-Nameki:2016cfr,Apruzzi:2016iac}.  The
theory on these D3-branes is interesting due to the variation of the
axio-dilaton -- and hence the D3-brane gauge coupling -- along the curve on
which the branes are wrapped. A description of the 2d $(0,2)$ theory resulting
from wrapped D3-branes in a 7-brane background has been achieved in
\cite{LSNW} with the help of a topological duality twist
\cite{Martucci:2014ema}, and corroborated considerably via duality with
M-theory.

{Apart from the 7- and D3-brane sectors, the supergravity sector forms the
  third defining constituent of} 2d $(0,2)$ F-theory compactifications.  The
gravitational sector of {\it perturbative} compactifications of Type IIB
string theory on Calabi--Yau fourfolds is well-known to give rise to a 2d
$(0,4)$ supergravity theory\cite{Dasgupta:1996yh,Gates:2000fj}, and some {further aspects of such perturbative 2d
  string compactifications have been investigated, for example, in
  \cite{Forste:1997bd,Font:2004et}. The gravitational multiplet gives rise to
  a so-called dilaton gravity theory as it comprises, apart from the 2d metric
  and the two gravitini, a real scalar field and its spin 1/2 superpartners,
  {the dilatini}. 

In this paper we will establish a description of the 2d $(0,2)$ supergravity
sector from F-theory {compactified on an elliptically fibered Calabi--Yau
fivefold $Y_5$}. Our first task is to derive the supergravity spectrum in
terms of 2d $(0,2)$ multiplets. After reviewing, in section \ref{sec:IIB}, the
structure of the 2d $(0,4)$ Type IIB supergravity of
\cite{Dasgupta:1996yh,Gates:2000fj} we perform a Type IIB orientifold
projection in section \ref{sec:IIBorienti} and uplift the resulting theory to
F-theory in section \ref{sec:GravAnF}. We will find that the chiral theory is
free of gravitational anomalies, as {it} must be, provided we take into
account in addition the duality twisted D3-brane sector and its contribution
to the anomalies found in \cite{LSNW}. This serves as a strong consistency
check.

In section \ref{sec_CS} we analyze the 2d $(0,2)$ supergravity from the
perspective of the dual $N=2$ Supersymmetric Quantum Mechanics (SQM) obtained
by compactifying M-theory on the same Calabi--Yau fivefold $Y_5$. An
interesting feature of supersymmetry in one dimension is known as
automorphic duality \cite{Gates:2002bc,Bellucci:2005xn}, which enables us to match
the supergravity spectrum to the $N=2$ SQM multiplets found by comparison
with the results of} \cite{Haupt:2008nu} for M-theory on Calabi--Yau fivefolds.
The $N=2$ SQM contains a rich sector of Chern-Simons like couplings which encode, apart from crucial information about the 7-brane gauge sector \cite{Schafer-Nameki:2016cfr}, the structure of gravitational anomalies of the dual F-theory.
 Unlike in higher dimensions
 \cite{Intriligator:1997pq,Aharony:1997bx,Grimm:2010ks,Bonetti:2011mw,Grimm:2011fx,Cvetic:2012xn,Bonetti:2013cza,Bonetti:2013ela},
 we find that the D3/M2-brane sector plays  a pivotal role in
 matching these two descriptions.
 
 A delicate peculiarity of 1d M-theory compactifications is the appearance of
 a tadpole constraint, which is the strong coupling analogue of an effect
 analyzed first in \cite{Ganor:1996xg,Dasgupta:1996yh} for Type IIA
 compactifications to two dimensions. The Type IIA vacuum necessarily includes
 a number of spacetime-filling strings to cancel a tadpole for the
 Neveu-Schwarz $B$-field. Upon T-dualising to Type IIB theory on a circle the
 resulting winding charge translates into a left-right asymmetry of  the
 Kaluza-Klein momentum of the chiral theory
 \cite{Ganor:1996xg,Dasgupta:1996yh}. We investigate this effect in the light
 of F/M-theory duality in section \ref{sec:vacenergy} and relate it to the
 vacuum momentum of the chiral 2d F-theory compactified on a circle. {This
 reveals a subtle relation between the F- and M-theory vacua special to two
 dimensions.}

%%%%%%%%%%%%%%%%%%%%%%%%%%%%%%%%%%%%%%%%%%%%%%%%%%
%%%%%%%%%%%%%%%%%%%%%%%%%%%%%%%%%%%%%%%%%%%%%%%%%%
\section{Type IIB String Theory on Calabi--Yau Fourfolds} \label{sec:IIB}
%%%%%%%%%%%%%%%%%%%%%%%%%%%%%%%%%%%%%%%%%%%%%%%%%%
%%%%%%%%%%%%%%%%%%%%%%%%%%%%%%%%%%%%%%%%%%%%%%%%%%

Our starting point is the supergravity obtained by compactifying Type IIB string theory on a Calabi--Yau fourfold $X_4$ to two spacetime dimensions.
This theory preserves $(0,4)$ supersymmetry, and we now {summarize its
most important properties from the discussion in} \cite{Dasgupta:1996yh,Gates:2000fj}.

 The massless spectrum is determined entirely in terms of the geometric data of $X_4$. 
 {The K\"ahler form $J$} as well as the Ramond-Ramond (RR) 2-form potentials $B_2$ and $C_2$ each contribute $h^{1,1}(X_4)$ real scalar fields. Another $2 h^{3,1}(X_4)$ real scalars originate from the complex structure deformations, together with 2 real scalars $C_0$ and $\varphi$ from the Type IIB axio-dilaton. 
A second source of scalars is the reduction of the self-dual  RR 4-form potential $C_4$: the modes of $C_4$ along the $b_4^+(X_4) = 2 + h^{2,2}_{+}(X_4)$ self-dual and $b_4^-(X_4) = 2 h^{3,1}(X_4) + h^{1,1}(X_4) - 1$ anti-self-dual 4-forms give rise to a corresponding number of anti-chiral and chiral real scalar fields, respectively. Note that 
\bea
b^+_4(X_4) =  b^-_4(X_4) + \tau(X_4) \,,
\eea 
where $\tau(X_4)$ is the Hirzebruch signature of $X_4$ and given by
\be\label{taudef}
\ba
 \tau(X_4)
 &= 48 + 2 h^{1,1}(X_4) + 2 h^{3,1}(X_4) - 2 h^{2,1}(X_4) \cr 
& = {1\over 45} \int_{X_4} 7 p_2(X_4) - p_1(X_4)^2  \geq 0 \,.
 \ea
 \ee
  For a collection of useful index theorems on Calabi--Yau fourfolds see, for
  example, \cite{Klemm:1996ts}.
  {The non-negativity of $\tau(X_4)$ means that} $b^-_4(X_4)  = 2 h^{3,1}(X_4) + h^{1,1}(X_4) - 1$ pairs of chiral and anti-chiral scalars combine into non-chiral real scalar fields. Together with the scalar modes described above, these form the bosonic components of $(0,4)$ superfields.

The  $2 h^{3,1}(X_4)$ real scalars from $C_4$ join the  $2 h^{3,1}(X_4)$ real modes from the complex structure deformations to form the scalar part of $h^{3,1}(X_4)$ complex structure $(0,4)$ hypermultiplets.
The remaining $h^{1,1}(X_4) - 1$ real scalars from $C_4$ combine with an equal number of K\"ahler moduli and 
 $2(h^{1,1}(X_4) - 1)$  real scalar fields  from the decomposition of $B_2$ and $C_2$ into $h^{1,1}(X_4) - 1$ $(0,4)$ chiral K\"ahler moduli multiplets.
 The reason why there only $h^{1,1}(X_4) - 1$, and not  $h^{1,1}(X_4)$, such moduli is because the real scalar field corresponding to the overall volume ${\cal V}(X_4)$ enters, along with the 2d metric $g^{(\rm 2d)}_{\mu \nu}$, the $(0,4)$ supergravity multiplet. 
 This leaves us with two more real scalar fields from the decomposition of $B_2$ and $C_2$ along this direction in moduli space, which join the two real scalars from the axio-dilaton to form the bosonic part of a universal (0,4) hypermultiplet.

%%%%%%%%%%%%%%%%%%%%%%%%%%%%%%%%%%%%%%%%%%%%%
%%%%%%%%%%%%%%%%%%%%%%%%%%%%%%%%%%%%%%%%%%%%%
\begin{table}\begin{center}
\begin{tabular}{|c|c|c|}\hline
$(0,4)$ Multiplet & Multiplicity& Bosonic Origin \cr\hline\hline
Hyper& $h^{1,1}(X_4) -1 $ & K\"ahler moduli, $B_2$, $C_2$, $C_4$ \cr \hline
Hyper & $ h^{3,1}(X_4)$ & Complex structure moduli, $C_4$ \cr \hline
Hyper  &1  & $C_0$, $\varphi$, $B_2$, $C_2$\cr \hline \hline
{Fermi}   & $ h^{2,1}(X_4)$  &  --\cr \hline \hline 
{Tensor} &  $\frac{1}{2} \tau(X_4)$ & $C_4$ \cr \hline\hline
Gravity & 1 & $g_{\mu \nu}$, ${\cal V}$\cr \hline
%  Fermis & $b_4^+(X_4)$ & $C_4$ \cr\hline
\end{tabular}
\caption{Spectrum of IIB on a Calabi--Yau fourfold $X_4$ in terms of the 2d $(0,4)$ multiplet structure.} \label{spectrumX4}
\end{center}
\end{table}
%%%%%%%%%%%%%%%%%%%%%%%%%%%%%%%%%%%%%%%%%%%%%
%%%%%%%%%%%%%%%%%%%%%%%%%%%%%%%%%%%%%%%%%%%%%

The remaining 
 \be
 n_-^{s}- n_+^{s} = b_4^+(X_4) - b_4^-(X_4) =  \tau(X_4)  
 \ee
anti-chiral real scalars {constitute the on-shell content of ${1\over 2}\tau(X_4)$ $(0,4)$ tensor multiplets. Massless fermions arise from} $n_f^\pm$ chiral and anti-chiral fermionic  Majorana-Weyl zero modes obtained by dimensional reduction of the two negative chirality Type IIB Majorana-Weyl gravitini $\Psi_{\mu, 1}^{-}$ and  $\Psi_{\mu,2}^{-}$. Their difference follows from an index theorem to be
\be\label{Gravitini}\ba
n_-^f - n_+^f  &=  2 \,  \hbox{ind} \slash{D}_{3/2} =  4 (-h^{1,1}(X_4) + h^{2,1}(X_4) - h^{3,1}(X_4))  = 2 \left(16 - {1\over 3 }\chi(X_4) \right)\,,
\ea\ee
where $\chi(X_4)$ denotes the Euler number of $X_4$.
By $(0,4)$ supersymmetry, the $n_+^f $ positive chirality Majorana-Weyl zero-modes form the fermionic part of the $(0,4)$ hypermultiplets whose bosonic components have been described already, with four Majorana-Weyl fermions per $(0,4)$ hypermultiplet. The number of such multiplets is  $h^{1,1}(X_4) + h^{3,1}(X_4)$, and hence 
\bea
 n_+^f  = 4(h^{1,1}(X_4) + h^{3,1}(X_4)), \qquad  n_-^f  = 4 h^{2,1}(X_4)\,.
 \eea
{The  $n_-^f$ anti-chiral real fermions from the gravitini contribute the equivalent of a total of $n_{\rm Fermi}$ $(0,4)$ Fermi multiplets, with 
\be
n_{\rm Fermi} 
=  h^{2,1}(X_4)\,.
\ee }
Finally, the 10d Type IIB dilatini $\lambda^+_1$ and $\lambda^+_2$ as well as the 10d gravitini $\Psi_{\mu,1}^{-}$ and  $\Psi_{\mu,2}^{-}$ each give rise to $ \hbox{ind} \slash{D}_{1/2}=2$ chiral Majorana-Weyl fermions and  $ \hbox{ind} \slash{D}_{1/2}=2$ anti-chiral Majorana Rarita-Schwinger modes, respectively. These are part of the gravity multiplet. This structure is summarized in table \ref{spectrumX4}. 

This chiral spectrum {has been shown in} \cite{Dasgupta:1996yh} to be free of 2d gravitational anomalies.
{For later reference, let us recapitulate the argument: }
In our normalization the contribution of {a real chiral scalar,}  a complex chiral Weyl fermion and a complex chiral spin 3/2 field to the gravitational anomaly polynomial in $d= 2k +2 = 2$ dimensions is given, respectively, by
\be
\ba
{I_{4, 0}} &= { - \frac{1}{8} L(T)|_{4- \rm form} = - \frac{1}{24} p_1(T) }\cr 
I_{4,1/2} &= \hat A(T)|_{4-\rm form} = - \frac{1}{24} p_1(T) \cr
I_{4,3/2} &= \Big(-1 + 2\sum_{i=1}^{2k+1}  \, {\rm cosh} (\lambda_i/2\pi)\Big) \hat A (  R ) |_{4-\rm form} = - \frac{1}{24} p_1(T) \times (-23) \,.
\ea
\ee
We refer to \cite{AlvarezGaume:1983ig} for details on the notation. 
The anomaly coefficient  $\mathfrak{a}$, defined via $I_4 = - \frac{1}{24} p_1(T) \, \mathfrak{a}$, of the discussed spectrum is therefore
\be
 \mathfrak{a} = \left(48 + (n^{s}_+ - n^{s}_-) + {1\over 2} (n_+^{f} - n_-^f) \right). \label{a04}
\ee
The overall $48 = 2 (1 + 23)$ is the contribution from the two complex chiral Weyl spin 1/2 and anti-chiral spin 3/2 fermions in the $(0,4)$ gravity multiplet.
The expression (\ref{a04}) indeed vanishes identically  \cite{Dasgupta:1996yh}.\footnote{Note that $\chi(X_4)= \int_{X_4} p_2(X_4)/2 - p_1(X_4)^2/8$ and the Dirac index for the dilatino is $\hbox{ind} \slash{D}_{1/2} = 2 = 1/1440 \int_{X_4} 7/4\,  p_1(X_4)^2 - p_2(X_4)$.}

%%%%%%%%%%%%%%%%%%%%%%%%%%%%%%%%%%%%%%%%%%%%%%%%%%
%%%%%%%%%%%%%%%%%%%%%%%%%%%%%%%%%%%%%%%%%%%%%%%%%%
\section{2d $(0,2)$ Type IIB Orientifolds}
\label{sec:IIBorienti}
%%%%%%%%%%%%%%%%%%%%%%%%%%%%%%%%%%%%%%%%%%%%%%%%%%
%%%%%%%%%%%%%%%%%%%%%%%%%%%%%%%%%%%%%%%%%%%%%%%%%%

Next we consider the orientifold action on the {spectrum of} {Type IIB
on Calabi--Yau fourfold}, with the goal to
obtain a $(0,2)$ supersymmetric theory in 2d. {The uplift of this to
F-theory will be  discussed in the next section}.

As usual in a Type IIB orientifold with projection $\Omega (-1)^{F_L} \sigma$ and O3/O7-planes, the K\"ahler form and holomorphic 4-form transform under the holomorphic involution $\sigma: X_4 \rightarrow X_4$ as 
\bea
\sigma: \qquad J \rightarrow J\,, \quad \Omega_{4,0} \rightarrow - \Omega_{4,0} \,,
\eea
while the form potentials transform under worldsheet parity as
\bea
\Omega (-1)^{F_L} : \qquad (C_0, B_2, C_2, C_4) \ \rightarrow \ (C_0, -B_2, -C_2, C_4)\,.
\eea
In particular, the massless scalar modes from invariant (anti-invariant) form potentials  are counted by the invariant (anti-invariant) eigenspace of the relevant cohomology groups.
It is therefore clear already from the bosonic spectrum that the $(h^{1,1}(X_4) -1)$ (0,4) K\"ahler hypermultiplets split up into $h^{1,1}_+(X_4) -1$ $(0,2)$ chiral K\"ahler multiplets containing the modes of $J$ and $C_4$  as well as  $h^{1,1}_-(X_4) $ axionic moduli containing the $B_2$ and $C_2$ modes. 
The universal $(0,4)$ hypermultiplet  reduces to a $(0,2)$ chiral multiplet
containing the axio-dilaton, while its $(B_2, C_2)$ components are projected
out. The overall volume ${\cal V}$ remains in the gravity multiplet of the
resulting $(0,2)$ dilaton gravity because it has no axionic partner with which to form a chiral multiplet.  
Finally, the complex structure moduli of the (0,4) parent theory descend to $h^{3,1}_-(X_4)$ $(0,2)$ chiral multiplets containing the complex structure moduli of the orientifold as well as  $h^{3,1}_+(X_4)$ $(0,2)$ chiral multiplets whose bosonic part is composed of the orientifold even $C_4$ modes.  

{The axion $C_4$ contributes $(0,2)$ tensor multiplets, consisting of one anti-chiral real scalar each. As $C_4$ is orientifold invariant, there are 
$\tau_+(X_4)$ such tensor multiplets, where $\tau_+$ is the orientifold-invariant part of the signature $\tau$.

The spectrum of positive chirality Weyl fermions follows these decompositions such as to form the fermionic part of the above $(0,2)$ chiral multiplets. In addition, we find $(0,2)$ Fermi multiplets, each containing one complex negative chirality Weyl spinor: Recall that in the $(0,4)$ theory there are $2 h^{2,1}(X_4)$ negative chirality Majorana Weyl fermions from each of the two 10d Majorana gravitini.}
Under the orientifold projection, the two 10d  gravitini are exchanged such that we can form one invariant and one anti-invariant combination.
The invariant combination contributes $2 h^{2,1}_+(X_4)$ Majorana-Weyl fermions, and the anti-invariant one $2 h^{2,1}_-(X_4)$ Majorana-Weyl fermions.
Altogether we hence find {$h^{2,1}_+(X_4) + h^{2,1}_-(X_4)$} $(0,2)$ Fermi multiplets.
Finally the gravity multiplet contains, in addition to the metric and the overall volume mode ${\cal V}$, one positive chirality complex dilatino $\lambda^+$ and one negative chirality gravitino $\psi_\mu^-$.  

The introduction of an orientifold enforces the presence of spacetime filling 7-branes and 3-branes due to tadpole constraints, and consequently introduces additional chiral and Fermi
multiplets. 
The simplest consistent 7-brane configuration consists of a single orientifold invariant 7-brane in the class $[S] = 8 [\rm O7]$ with  $[\rm O7]$ denoting the divisor class wrapped by the O7-plane. In this case the 7-brane gauge group is trivial, but the pure 7-brane sector  contributes 2d $(0,2)$ multiplets from the brane moduli sector. 
Since $S$ is singular, this spectrum is most conveniently counted in terms of cohomology groups of its normalization $\hat S$ (see \cite{Collinucci:2008pf} for a discussion for Type IIB orientifolds on Calabi--Yau threefolds).
We do not present a full derivation of this moduli sector here, but note that in analogy with compactifications to four dimensions \cite{Jockers:2004yj} one expects 
$h^{1,0}_-(\hat S)$ chiral multiplets containing the Wilson line moduli and $h^{3,0}_-(\hat S)$ chiral multiplets associated with the brane deformation moduli.
Furthermore in analogy with the explicit analysis \cite{Schafer-Nameki:2016cfr,Apruzzi:2016iac} of the 7-brane spectrum in 2d F-theory  there are   $h^{2,0}_-(\hat S)$
 extra Fermi multiplets. In situations with non-trivial gauge groups  
all of these multiplets transform in the adjoint representation, but for trivial gauge group they are counted as part of the open-closed moduli space and hence become part of the supergravity moduli sector in F-theory.  These results are summarized in the second column of table \ref{tab:SpecO}.
The D3-spectrum introduces extra degrees of freedom which are the type IIB analogue of the spectrum derived in detail in \cite{LSNW} in F-theory language, to which we refer for more details on this sector.

\section{ F-theory Uplift  and Gravitational Anomaly Cancellation} \label{sec:GravAnF}

Having established the {$N=(0,2)$ spectrum for the Type IIB orientifold compactification on a Calabi--Yau fourfold, }
we can now uplift this spectrum to F-theory on a Calabi--Yau fivefold $Y_5$, which is elliptically fibered over a K\"ahler base $B_4$. 
In the orientifold limit, the IIB fourfold $X_4$ is the double cover of $B_4$, but the spectrum which we will find is valid for general F-theory backgrounds. 
{Indeed, deforming away from the orientifold limit corresponds to motion in the
complex structure moduli space of the elliptic fibration, and as long as such
a deformation does not render the fibration singular, related to the
introduction of 7-branes, no new massless degrees of freedom are
introduced.}

The uplift of the cohomology groups proceeds according to the familiar rules for comparing the invariant and anti-invariant cohomology groups on $X_4$ to the cohomologies on the elliptic fibration $Y_5$. These rules have been established for F-theory on elliptic fourfolds e.g. in \cite{Blumenhagen:2010ja,Clingher:2012rg}. 
The invariant cohomology groups of $X_4$ lift to cohomology groups on $B_4$ such that  
\be
h^{p,q}_+ (X_4) \ \rightarrow \ h^{p,q} (B_4) \,,
\ee
whereas the lift of the anti-invariant cohomology groups contributes as
\be \label{uplift1}
h^{p,q}_- (X_4) \ \rightarrow \ h^{p+1,q}(Y_5) - h^{p+1 , q}(B_4) \,.
\ee
This can be thought of  as the lift of an anti-invariant $(p,q)$ form to a $(p+1,q)$ form on $Y_5$  with one extra leg in the elliptic fiber. The idea is that the monodromy in the fiber as one encircles the O7-plane compensates for the fact that the base component of the form is anti-invariant.
The F-theory moduli spectrum in addition encompasses the chiral and Fermi multiplets counted by the cohomology groups $H^{i,0}_-(\hat S)$, $i=1,2,3$ with $\hat S$ the normalization of the divisor $S$ wrapped by a {\it single} 7-brane (with no gauge group).
These cohomology groups feed into the cohomology of the F-theory uplift as follows
\bea\label{uplift2}
h^{i,0}_-(\hat S) \ \rightarrow \  h^{i+1,1}(Y_5) - h^{i+1 , 1}(B_4) \,.
\eea
Both (\ref{uplift1}) and (\ref{uplift2}) can be justified more formally in terms of a Leray spectral sequence applied to a suitable stable degeneration limit which is the mathematical incarnation of the perturbative Sen limit of F-theory \cite{Clingher:2012rg}.  
Applying these maps to the cohomology groups counting the Type IIB orientifold matter leads to the corresponding matter in F-theory  as displayed in the third column of table \ref{tab:SpecO}.

%%%%%%%%%%%%%%%%%%%%%%%%%%%%%%%%%%%%%%%%%%%%%
%%%%%%%%%%%%%%%%%%%%%%%%%%%%%%%%%%%%%%%%%%%%%
 \begin{table}\begin{center}
\begin{tabular}{|c||c|c||c||c|}\hline
\multirow{2}{*}{$(0,2)$ Multiplet} & \multirow{2}{*}{IIB Orientifold} &\multirow{2}{*}{F-theory} & Origin& \multirow{2}{*}{SQM Multiplet} \cr
 & & & in IIB/F-theory & \cr\hline\hline
Chiral & $h^{1,1}_+(X_4) -1$ & $h^{1,1}(B_4) -1$    &$J,C_4$  & $(1,2,1)$ \cr \hline
 & $h^{1,1}_-(X_4)$ & \multirow{2}{*}{$h^{2,1}(Y_5) - h^{2,1}(B_4)$}  &$B_2, C_2$  & \multirow{2}{*}{$(2,2,0)$} \cr 
 &   $h^{1,0}_-(\hat S)$ &        & Wilson lines &  \cr \hline
  &1  & \multirow{3}{*}{$h^{4,1}(Y_5) $}   &$C_0, \varphi$ &\multirow{3}{*}{$(2,2,0)$} \cr 
  & $h^{3,1}_-(X_4)$ & &{\rm cmplx. str.} &\cr
  & $h^{3,0}_-(\hat S)$ & & brane def. &\cr \hline
   & $h^{3,1}_+(X_4)$&  $h^{3,1}(B_4)$  &$C_4$  & $(0,2,2)$ \cr \hline\hline
{Tensor} &$\tau_+(X_4)$ & $\tau(B_4)$  &$ C_4$ &   $(0,2,2)$  \cr \hline\hline
 Fermi & $h^{2,1}_+(X_4)$ & $h^{2,1}(B_4)$  &$  - $& $(2,2,0)$  \cr \hline
   &  $h^{2,1}_-(X_4)$  & \multirow{2}{*}{$h^{3,1}(Y_5) - h^{3,1}(B_4)$} &$-$  &   \multirow{2}{*}{$(0,2,2)$}   \cr
   &  $h^{2,0}_-(\hat S)$ &   & $-$& \cr \hline\hline
   \multirow{2}{*}{Gravity} & \multirow{2}{*}{1} & \multirow{2}{*}{1}
   &\multirow{2}{*}{$g_{\mu\nu}, {\cal V}$} &  $(1,2,1)$ \cr 
   & & & & + 1d gravity\cr \hline 
\end{tabular}
\caption{Moduli and gravitational sector of IIB orientifolds on Calabi--Yau fourfold
  $X_4$ as well as of F-theory on $Y_5$ in terms of 2d $(0,2)$ multiplets. The
  {subscripted} signs in the second column indicate the orientifold even and
  odd parts of the cohomology. 
  {The fourth column contains the origin of the modes from the bosonic sector of IIB/F-theory.}
  The last column indicates the dual 1d $N=2$ SQM multiplets
  obtained from M-theory on the same $Y_5.$\label{tab:SpecO}}
\end{center}
\end{table}
%%%%%%%%%%%%%%%%%%%%%%%%%%%%%%%%%%%%%%%%%%%%%
%%%%%%%%%%%%%%%%%%%%%%%%%%%%%%%%%%%%%%%%%%%%%

A number of remarks are in order concerning the F-theory spectrum:
%First, $h^{4,1}(Y_5) - h^{4,1}(B_4) = h^{4,1}(Y_5)$ (since $h^{4,1}(B_4)=0$ for dimensional reasons of course) counts both the complex structure moduli of the Type IIB orientifold, the axio-dilaton and in addition the 7-brane moduli which are necessarily present.
If the elliptic fibration $Y_5$ is smooth the gauge group from the 7-branes is trivial and no charged matter arises from the pure 7-brane sector in addition to the moduli fields listed in table \ref{tab:SpecO}. 
For non-abelian stacks of 7-branes the matter states propagating along the 7-brane divisors (`bulk matter') transform in the adjoint of the non-abelian gauge group (or its unbroken subgroups in the presence of gauge flux). 
Matter localised at the intersection of 7-branes contributes further $(0,2)$ chiral and Fermi multiplets to the spectrum. Such charged matter has already been discussed in detail  in \cite{Schafer-Nameki:2016cfr, Apruzzi:2016iac}, to which we refer for more details. 
In this case, one finds also  
\begin{equation}
  N_{\rm vectors} = h^{1,1} (Y_5) - h^{1,1} (B_4) -1 \,, 
\end{equation} 
vector multiplets from the 7-branes, where each of the vectors contains a single negative chirality complex fermion. 
If we restrict ourselves to smooth elliptic fibrations for simplicity, none of these extra multiplets arise.

However, even for smooth Calabi--Yau fivefolds a D3-tadpole constraint enforces the presence of spactime-filling D3-branes wrapping suitable curves on $B_4$. 
The total curve class $C$ wrapped by such D3-branes  can be determined via duality with M-theory as discussed in more detail in section \ref{sec_CS}.
The spectrum on these D3-branes has been investigated in detail in \cite{LSNW} and is summarized in table \ref{tab:D3spec}.

 \begin{table}\begin{center}
\begin{tabular}{|c|| c | c|}\hline
$(0,2)$ Multiplet &   Multiplicity & Origin   \cr\hline\hline
Chiral & $h^{0}(C,N_{C/B_4})$ & Deformations  \cr\hline
          &  $g - 1 + c_1(B_4) \cdot C$ & Wilson lines \cr \hline\hline
Fermi &  $     h^{0}(C,N_{C/B_4}) + g - 1 - c_1(B_4) \cdot C$ & D3-fermions \cr \hline      
           &   $8 c_1(B_4) \cdot C$      & 3-7 strings \cr\hline
\end{tabular}
\caption{Matter spectrum on a D3-brane on curve $C$ on $B_4$ of genus $g$. \label{tab:D3spec}}
\end{center}
\end{table}

We are now in a position to compute the gravitational anomaly of the 2d $(0,2)$ theory obtained from F-theory on $Y_5$, where for simplicity we focus on a smooth fibration with trivial 7-brane gauge group.  
Summing up the contribution of  the chiral, Fermi {and tensor} multiplets in the moduli sector, listed in table \ref{tab:SpecO}, to the gravitational anomaly polynomial yields
\bea \label{I4moduli}
I_{4, \rm moduli} = - \frac{1}{24} p_1(T)  \Big( - \tau(B_4) + \chi_1(Y_5) - 2 \chi_1(B_4) \Big) = - \frac{1}{24} p_1(T) \,   {\mathfrak a}_{\rm moduli}\,.
\eea 
Here we have combined the Hodge numbers in table \ref{tab:SpecO}, weighted by signs {$+ (-)$} for the chiral ({Fermi and tensor}) multiplets, respectively, into the topological indices 
\be
\chi_q(M) = \sum_{p=1}^{{\rm dim}(M)} (-1)^p h^{p,q} (M) \,.
\ee
This uses in particular that, since $Y_5$ is a smooth model with trivial 7-brane gauge group, by the Tate-Shioda-Wazir theorem  $h^{1,1}(Y_5) - h^{1,1}(B_4) - 1=0$.
The D3-brane sector studied in \cite{LSNW} contributes furthermore 
with $c_L - c_R = 6 c_1 (B) \cdot C$, as follows from table \ref{tab:D3spec},
\be \label{I4D3}
I_{4, \rm D3} = - \frac{1}{24} p_1(T) \Big(  -  6 c_1(B_4) \cdot C \Big) = - \frac{1}{24} p_1(T)  \,  {\mathfrak a}_{\rm D3} \,.
\ee
The class $C$ of the curve wrapped by  the D3-branes is fixed by a tadpole cancellation condition which will be discussed in more detail in section \ref{sec_CS}.
In particular, for an F-theory compactification on a smooth elliptic fibration (in absence of non-abelian gauge groups from the 7-brane sector), the class of $C$ is fixed as
\bea \label{Csmooth}
[C] =  15 \, c_1^3 + \frac{1}{2} \, c_1 \, c_2 \,,
\eea
where $c_i \equiv c_i(B_4)$.
In addition, the gravity multiplet yields an anomaly of
\be
I_{4, \rm grav} = - \frac{1}{24} p_1(T) \Big(  1 + 23  \Big) = - \frac{1}{24} p_1(T)  \,  {\mathfrak a}_{\rm grav} \,,
\ee
where the two terms are due to the positive chirality dilatino and the negative chirality gravitino respectively.  
This results in a total gravitational anomaly of the form
\be
I_{4}  =  -{1\over 24} p_1(T) \cdot \Big(
 -\tau(B_4)  + \chi_1(Y_5) - 2 \chi_1(B_4)+ 24 + \mathfrak{a}_{D3} \Big) \,.
\ee
The combined expression can be written in terms of the Chern
classes $c_i(B_4) \equiv c_i$ of the base fourfold $B_4$ with the help of the identities
(see e.g. \cite{Klemm:1996ts, Haupt:2008nu}) 
\be
\ba
\chi_0 (B_4) 
&={1\over 720}
\int_{B_4}(-c_4 + c_1c_3 + 3c_2^2 + 4c_1^2 c_2 - c_1^4)\cr
\chi_1 (B_4) 
&={1\over 180}
\int_{B_4}(- 31 c_4 -14 c_1c_3 + 3c_2^2 + 4c_1^2 c_2 - c_1^4)\cr
\chi_1(Y_5) 
&= -{1\over 24} \int_{Y_5} c_5 (Y_5)  
= \int_{B_4} (90 c_1^4 + 3 c_1^2 c_2 - {1\over 2} c_1 c_3) \cr
\tau(B_4) & = %= {\chi \over 3} + 32 = 2 (8 + h^{1,1} + h^{3,1} - h^{2,1}) + 32  
% \cr &=
 {1 \over 180}\int_{B_4} (12 c_2^2 - 56 c_1 c_3 + 56 c_4 - 4 c_1^4 + 16 c_1^2
 c_2) 
\,.
\ea
\ee
In the third line we have used adjunction to express $c_5(Y_5)$ in terms of $c_i$ for a smooth Calabi--Yau elliptic fibration over $B_4$. 
The anomaly can then be simplfied to 
\be
I_4 =  -{1\over 24} p_1(R) \Big(-24 \chi_0(B_4) + 24\Big) \,.
\ee
This uses in addition the expression (\ref{Csmooth}) appropriate for F-theory on smooth Weierstrass models. 
Since we require that $B_4$ provides the base space of an elliptically fibered
Calabi--Yau fivefold it is inconsistent for $B_4$ to have any
$(0,4)$-forms, as these would uplift to destroy the Calabi--Yau property of
$Y_5$. This, combined with the K\"ahler nature of $B_4$, fixes $h^{0,k}(B_4) = 0$ for
all non-zero $k$, and $h^{0,0}(B_4) = 1$. Thus 
\begin{equation}
  \chi_0 (B_4) = 1 \,,
\end{equation}
and the anomaly is cancelled automatically for all smooth elliptic fibrations. 

This result generalizes in the presence of non-abelian singularities in the fiber, associated with non-trivial gauge groups 
on the 7-branes. In this situation, as stressed before, charged massless
matter arises along the 7-branes in the adjoint representation of the gauge
group as well as localised matter at the intersection of the 7-branes
\cite{Schafer-Nameki:2016cfr,Apruzzi:2016iac}. The representation of the
latter type of massless states depends on the specifics of the setup. The
contribution of this charged matter to the gauge anomalies has been detailed
in \cite{Schafer-Nameki:2016cfr}, and similarly its contribution to the gravitational anomaly can be expressed in terms of topological indices involving the divisor classes of the wrapped 7-branes on the base.
At the same time, the expression (\ref{Csmooth}) for the curve class of the wrapped D3-branes picks up extra terms proportional to that same 7-brane divisor classes. This is because, as will be discussed in the next section, (\ref{Csmooth}) results from the explicit form of 
$c_4(Y_5)$, which in turn changes in the presence of 7-branes carrying gauge groups because the singular fibration must first be resolved. These two additional contributions cancel each other off as can be checked in analogy to the examples studied in \cite{Schafer-Nameki:2016cfr}. We note, however, that the contribution to the gravitational anomalies from the D3-brane sector as a function of $C$ is unchanged compared to the smooth case: This sector universally contributes $- 6 c_1(B_4) \cdot C$ to the anomaly polynomial \cite{LSNW}.

%%%%%%%%%%%%%%%%%%%%%%%%%%%%%%%%%%%%%%%%%%%%%%%%%

\section{Match with $N=2$ SQM from Dual M-theory}\label{sec_CS}

The 2d $(0,2)$ theory obtained by compactifying F-theory on $Y_5$ is related, via compactification on a circle $S^1$, to an $N=2$ Supersymmetric Quantum Mechanics (SQM) arising as the low-energy effective theory of M-theory on the same $Y_5$.
The $N=2$ SQM resulting from M-theory on a general Calabi--Yau fivefold has been worked out in \cite{Haupt:2008nu}. 

Let us first compare the F-theory spectrum found in table  \ref{tab:SpecO} to the M-theory results.
A priori a 2d $(0,2)$ chiral multiplet descends, upon $S^1$ reduction, to a $(2,2,0)$ multiplet in the SQM, where the notation $(n,2,2-n)$ refers to a multiplet with $n$ real scalar fields, $2$ real fermion fields and $2-n$ real auxiliary fields. 
Similarly, a 2d $(0,2)$ Fermi multiplet reduces at first sight to a $(0,2,2)$ multiplet. 
However, a property special to 1d supersymmetric theories  known as automorphic duality \cite{Gates:2002bc}  allows us to dualise real scalar fields into auxiliary fields provided the moduli space metric does not depend on them \cite{Bellucci:2005xn}. This is the case in particular for axionic pseudo-scalars which arise from reduction of the Type IIB RR or NS-NS forms and hence enjoy a shift symmetry\footnote{{In principle, the perturbative shift symmetry can be broken by instanton effects. The observed match between the M/F-theory spectra suggests that the automorphic duality survives such instanton effects in the present case. It would indeed be interesting to understand this in more detail.}}. 
This way for instance the $(2,2,0)$ multiplets associated with the K\"ahler multiplets can be dualised into $(1,2,1)$ multiplets because they contain a real axionic field from reduction of $C_4$. In the last column of table \ref{tab:SpecO} we are collecting the $N=2$ SQM multiplets obtained by suitable dualisation corresponding to the multiplets one finds  \cite{Haupt:2008nu}  from dimensional reduction of M-theory on $Y_5$. To stay in the example of the K\"ahler moduli, the auxiliary field is given by the 1-form obtained from reduction of the M-theory 3-form $C_3$ along the associated 2-cycles; since the 1-form field in the massless multiplet is harmonic it can be written as a total time derivative along $\mathbb R$, which is precisely the form of an auxiliary field in SQM.
 
The other notable application of automorphic duality here concerns the $h^{2,1}(B_4)$ multiplets, which arise as 2d $(0,2)$ Fermi multiplets in F-theory.
In the dual M-theory picture these are obtained as $(2,2,0)$ SQM multiplets with the two real scalars per multiplet  corresponding to the modes of the M-theory 3-form expanded in terms of the base $(2,1)$ and $(1,2)$ forms. Since the moduli space metric, derived in \cite{Haupt:2008nu}, does not depend on these scalars, they can both be dualised into auxiliary fields such as to conform with the $S^1$ reduction of the F-theory Fermi multiplet. Finally the 2d gravity multiplet splits up into a (1,2,1) multiplet containing the real scalar ${\cal V}$ and the dilatino, as well as a 1d 'gravity' multiplet containing a 1d lapse function as well as the gravitino mode \cite{Haupt:2008nu}. In summary, a perfect match between the F-theory and the dual M-theory spectra has been achieved.

Of special importance are the 
Chern-Simons (CS) couplings of the dual $N=2$ SQM,  which take the form \cite{Haupt:2008nu}
\bea \label{CSdef1}
S_{\rm CS} = 2 \pi \sum_I  (k^I_{\rm curv} - k^I_{\rm M2})  \int_{\mathbb R} A_I  \,.
\eea
The 1-form potentials $A_I$ are precisely the aforementioned top-forms in 1d
which arise by reducing the M-theory 3-form as $C_3 = \sum_I A_I \cdot D_I$
with $D_I \in H^{1,1}(Y_5)$. The two couplings in (\ref{CSdef1}) originate
from the M-theory CS coupling  $\int_{\mathbb R^{1,10}} C_3 \cdot X_8$ and
the coupling to M2-branes.
In absence of 4-form flux, which we assume for simplicity, the couplings are \cite{Haupt:2008nu}
\bea
k^I_{\rm curv} = \frac{1}{24} \int_{Y_5 }D_I \cdot c_4(Y_5), \qquad k^I_{\rm M2} = \int_{Y_5 } D_I \cdot [{\rm M2}] \,.
\eea
Tadpole cancellation enforces 
\bea \label{tadpoleM2}
[{\rm M2}] = \frac{1}{24} c_4(Y_5)
\eea
 so that the net coupling $S_{CS}$ vanishes identically. Nevertheless the two individual terms in (\ref{CSdef1}) contain rich information about the spectrum.
Indeed, if the SQM is interpreted as the circle reduction of a 2d $(0,2)$ F-theory compactifications, then the CS couplings  are generated by integrating out the massive Kaluza-Klein (KK) states. The situation is similar to the generation of 5d or 3d CS terms by circle reduction from 6d or, respectively, 4d F-theory vacua \cite{Intriligator:1997pq,Aharony:1997bx,Grimm:2010ks,Bonetti:2011mw,Grimm:2011fx,Cvetic:2012xn,Bonetti:2013cza,Bonetti:2013ela}, and has already been analyzed in \cite{Schafer-Nameki:2016cfr} for 
the CS couplings associated with the 1-forms $A_i$ uplifting to $U(1)$ gauge fields in 2d.

In what follows we investigate the interplay between the 2d matter spectrum and the CS terms in 1d involving the 1d 1-form $A_0$, which unlike the gauge potentials $A_i$, is associated to the KK $U(1)$  arising in 1d as we  compactify the 2d $(0,2)$ theory on a circle. Unlike in higher dimensions, we will see that the M2-brane couplings play a crucial role in matching the 1d CS terms to loops from the 2d theory on a circle.
For simplicity we consider again an elliptic fibration without non-abelian fiber singularities, which can be described by a smooth Weierstrass model.
Denoting the holomorphic zero-section by $S_0$\footnote{{As usual the zero-section gives a copy of the base and intersects the fiber precisely once.}}, the curvature dependent CS couplings in 1d involve \be
\frac{1}{24} c_4(Y_5) =  S_0 \cdot \tW_3 + \tW_4 {= (S_0 +c_1) \cdot \tW_3 + (\tW_4 - c_1 \cdot \tW_3)}
\ee
with 
\be
 \tW_3 = 15 \, c_1^3 + \frac{1}{2} \, c_1 \, c_2, \qquad     \tW_4= 15 c_1^4 + \frac{1}{2} c_1^2 c_2 - \frac{1}{24} c_1 c_3 + \frac{1}{24} c_4.
\ee
{Note that $\frac{1}{24} c_4(Y_5)$ denotes a curve class on $Y_5$. Its component along the fiber is given by $S_0 \cdot \frac{1}{24} c_4(Y_5)$ times the fiber class {because the zero-section intersects the fiber precisely once}. Since  $S_0 \cdot (S_0 + c_1) =0$, the curve class in the fiber is therefore given by $\tW_4 - c_1 \cdot \tW_3$, while $ \tW_3$ denotes the {component of the} curve class {projected} on the base.}
Similarly we can expand 
\be
{[{\rm M2}] = (S_0 + c_1) \cdot C  + [{\rm M2}_F]\,,}
\ee
with $C$ and $[{\rm M2}_F]$ denoting the total class dual to the wrapped M2-branes on the base and along the elliptic fiber, respectively. Tadpole cancellation (\ref{tadpoleM2}) then requires 
\bea \label{TadpoleM2}
C= \tW_3  \qquad \quad [{\rm M2}_F] = {\tW_4 - c_1 \cdot \tW_3} \,.
\eea 
%{Note that $S_0 \cdot (S_0 + c_1) =0$ and therefore the curve class in the fiber is indeed given by $\tW_4 - c_1 \cdot \tW_3$.}
 The first condition in (\ref{TadpoleM2}) is the M-theory dual of the D3-brane tadpole cancellation condition and fixes the class of the M2/D3-brane in $B_4$.
 The second condition, on the other hand, is a consistency condition in M-theory which has no direct analog in the 2d F-theory vacuum. As stressed already in \cite{Schafer-Nameki:2016cfr}, it is the M-theory version of the tadpole condition \cite{Dasgupta:1996yh} enforcing the presence of a number of spacetime-filling fundamental strings in Type IIA compactifications on Calabi--Yau 4-folds. We will come back to this interpretation in section \ref{sec:vacenergy}. In order for the M-theory compactification to exist, this condition must be assumed to hold. Thus the CS couplings (\ref{CSdef1}) do not contain any terms proportional to the class {$\tW_4 - c_1 \cdot\tW_3$}, which cancels by assumption in the contributions from the curvature and the M2-brane. The remaining CS couplings take the form
 \bea
 k^0_{\rm curv} = \int_{Y_5} D_0 \cdot S_0 \cdot \tW_3, \qquad k^0_{\rm M2} = \int_{Y_5} D_0 \cdot S_0 \cdot C \,.
\eea
Of course, due to (\ref{TadpoleM2}), also these couplings eventually cancel in $S_{\rm CS}$, but we will treat $C$ as an independent class for the time being. Furthermore, we make the ansatz\footnote{{As usual, determining the generator, in terms of a divisor class on the elliptic fibration, of the Kaluza-Klein $U(1)$ in going from F-theory to M-theory by circle reduction requires a careful match of the respective effective field theories  \cite{Cvetic:2012xn}. Our arguments below show that the choice made here is indeed consistent.    }}
\bea \label{D0def}
D_0 = - S_0 - \alpha \, c_1 \,,
\eea
for the generator of the KK $U(1)$ field in the expansion $C_3 = A_0 \cdot D_0
+ \cdots$. Using the relation $S_0^2 = - S_0 \cdot c_1$ on the Weierstrass model of $Y_5$, the CS couplings take the form
\bea \label{kCS2}
k^0_{\rm curv}  = \frac{1 - \alpha }{24} \int_{B_4} c_1 \cdot (360 \, c_1^3 + 12 \, c_1 \, c_2)\,, \qquad  k^0_{\rm M2}  = (1 - \alpha) \int_{B_4} c_1 \cdot C.
\eea

These have to match the 1-loop induced couplings obtained by compactifying the 2d F-theory on an $S^1$ and integrating out the massive KK states.
The 1-loop couplings receive contributions from the spin $\frac{1}{2}$ fields contained in the moduli sector and the D3-brane sector, as well as from the spin $\frac{1}{2}$ and  $\frac{3}{2}$ fields in the $(0,2)$ gravity multiplet. 

Upon $S^1$ reduction each fermion gives rise to a tower of KK modes, carrying KK $U(1)$ charge $q_n =n$ and mass $m_n$. 
The loop induced contribution of a single 2d complex Weyl fermion of chirality $P = \pm 1$ to the 1d CS coupling of the KK $U(1)$ is then 
\be
\ba
\delta k^0_{{\rm 1-loop},\pm \frac{1}{2}} &=  - \frac{1}{2} P  \sum_{n=-\infty}^{\infty} q_n \, {\rm sign}(m_n)  =  - \frac{1}{2} P  \sum_{n=-\infty}^{\infty} n \, {\rm sign}(n) \cr
&  = - \frac{1}{2} P \, \times \,  2 \sum_{n=0}^{\infty} n    = -\frac{1}{24} (-2 P). 
\ea
\ee
In summing over the tower of KK states we are following the convention for the mass that $ {\rm sign}(n)  =  {\rm sign}(m_n)$.
This convention is responsible for the minus sign in the normalization of $D_0$ in (\ref{D0def}).\footnote{\label{Footnote5}More precisely, the relative signs between the CS terms in 1d and the 1-loop terms are chosen as in \cite{Schafer-Nameki:2016cfr}. Note that  \cite{Schafer-Nameki:2016cfr} considers  the $U(1)_i$ gauge groups associated with the exceptional divisors $D_i$ appearing in the resolution of a singular Weierstrass model. 
The $D_i$ contain a rational curve $F_i$ in the fiber, and an M2-brane wrapping $F_i$ has negative charge $q_i = D_i \cdot F_i = -1$ under $U(1)_i$.
By contrast, the zero-section $S_0$ intersects the generic fiber $ {\frak F} $ with positive intersection number,
$S_0 \cdot {\frak F} = 1$. The fiber ${\frak F}$ is the holomorphic curve wrapped by an M2 brane corresponding to one of the KK states. Hence we must include a minus sign in the definition of $D_0$ if we stick to the sign conventions of \cite{Schafer-Nameki:2016cfr}. } Furthermore we have used $\zeta$-function regularization $\sum_{n=1}^\infty n = \zeta(-1) =  - \frac{1}{12}$ as e.g. in \cite{Bonetti:2013ela} in 6d.
Summing over all fermions gives the contribution
\bea
k^0_{{\rm 1-loop},\pm \frac{1}{2}}  =  -\frac{2}{24} (n_- - n_+) =  - \frac{1}{24} \left(   - 2 ({\mathfrak a}_{\rm moduli} +{\mathfrak a}_{\rm D3} + 1 ) \right) 
\eea
with ${\mathfrak a}_{\rm moduli}$ and ${\mathfrak a}_{\rm D3}$   the contribution of the moduli and D3 sector to the gravitational anomaly polynomial
as computed in (\ref{I4moduli}) and (\ref{I4D3}), and the extra $+1$ denotes the contribution from the positive chiral Weyl fermion in the gravity multiplet. 
Similarly the contribution of the spin $\frac{3}{2}$ field in the gravity multiplet is 
\bea
k^0_{{\rm 1-loop},\pm \frac{3}{2}}  =  - \frac{1}{24} \left(   - 2 {\mathfrak a}_{3/2}  \right) =   - \frac{1}{24} \left(   - 2 \times 23  \right)  \,.
\eea

These terms match the CS couplings as follows: 
First, it is clear that the contribution from the D3 spectrum reproduces by itself the CS coupling due to the M2-brane because these are the only terms depending on the M2/D3-brane curve class $C$. Taking into account the sign in the definition  (\ref{CSdef1}) of the couplings, this means that $ - k_{\rm M2}^0 = k^0_{\rm 1-loop,D3}$.
This allows us to fix the parameter $\alpha$ in (\ref{D0def}) by comparing
(with {${\mathfrak a}_{\rm D3} = - 6 c_1(B_4) \cdot C$}) 
\bea
 - k_{\rm M2}^0 =  - (1-\alpha) \int_{B_4} C \cdot c_1, \qquad \quad k^0_{\rm
 1-loop,D3}  =      \frac{2}{24}     {\mathfrak a}_{\rm D3} = -  \frac{1}{2}
 \int_{B_4} C \cdot c_1 \,. 
\eea
Hence
\be\label{HalfIt}
D_0 = - S_0 - \frac{1}{2} c_1(B) \,.
\ee
This is the precise two-dimensional analog of how the coefficient $\alpha$ is fixed in higher dimensions, here by requiring the absence of the CS coupling for $A_0$:
Absence of this CS coupling is equivalent to tadpole cancellation, which fixes $[{\rm M2}] = C$. Furthermore, consistency with F-theory implies  $ - k_{\rm M2}^0 = k^0_{\rm 1-loop,D3}$, which eventually fixes $\alpha$.
Up to the overall sign, which is explained in footnote \ref{Footnote5}, this matches precisely with the expression for $D_0$ in \cite{Park:2011ji} in 6d F-theory compactifications. There $\alpha$ was fixed to this value by requiring that the CS coupling $A_\beta \wedge F_0 \wedge F_0$ associated to the KK $U(1)$, which has a coefficient proportional to $D_0 \cdot D_0 \cdot D_\beta$, has to vanish. Here $D_\beta$ are divisors pulled back from the base and $A_\beta$ the associated gauge field. This results in the linear combination (\ref{HalfIt}). 

With this normalization, inspection of (\ref{kCS2}) and of  ${\mathfrak a}_{\rm moduli}$ and $ {\mathfrak a}_{3/2}$ entering the 1-loop contributions  shows that indeed
\bea
k_{\rm curv}^0 = k_{\rm 1-loop, moduli}^0 + k_{\rm 1-loop, 3/2}^0 \,.
\eea
In particular this explains why the D3 tadpole cancellation condition  in (\ref{TadpoleM2}) is necessary and sufficient for the cancellation of the 2d gravitational anomaly since the vanishing of the CS terms in M-theory is equivalent to the cancellation of the 2d gravitational anomalies between the moduli and gravity sector on the one hand and the D3-brane sector on the other hand.

\section{M-theory Tadpoles in Light of F/M-theory Duality}\label{sec:vacenergy}

As pointed out in the previous section, the M-theory tadpole condition ${\rm [M2]} = \frac{1}{24} c_4(Y_5)$ with $\frac{1}{24} \,c_4(Y_5) = S_0 \cdot {\tW}_3 + \tW_4$ requires the introduction of M2 branes wrapping a curve in the class $[C] =  \tW_3$ in the base. This is the M-theory dual of the D3-brane tadpole constraint in F-theory. But in addition it enforces the presence of a number of 
\bea\label{labelnF}
n_F = \int_{B_4}{\tW}_4  {- c_1 \cdot \tW_3} \,,
\eea
M2-branes wrapping the elliptic fiber in M-theory. This constraint has no
analogue in the dual F-theory. Indeed, already for an $N = (2,2)$
supersymmetric  2d Type IIA string compactification on a Calabi--Yau fourfold
$X_4$ the 10d CS coupling $\int_{\mathbb R^{1,9}} B_2 \cdot X_8$ requires
introducing a number of $\chi(X_4)/24$ spacetime-filling fundamental strings
\cite{Sethi:1996es}. The M2-branes wrapped along the elliptic fiber are the
M-theory incarnation of these fundamental strings.  In this section we will
interpret the effect of these M2-branes in the light of F/M-theory duality.
Recall that the M-theory SQM is the $S^1$ reduction of a dual 2d $(0,2)$
F-theory upon performing a T-duality along that circle. The $S^1$ is
identified with one of the circles in the elliptic fibration, and so the
wrapped M2-branes carry winding modes along the $S^1$.  A similar observation
has been made about a 2d Type IIA vacuum: After reducing the vacuum along an  $S^1$,
the wound strings required by the tadpole carry non-trivial winding number \cite{Ganor:1996xg,Dasgupta:1996yh,Gates:2000fj}. After T-dualising along the
$S^1$ the winding modes translate into momentum modes.  The latter are in fact
interpreted as the momentum modes of the massless fields of the 2d $(0,2)$
vacuum compactified on $S^1$, and the total number of winding modes must match
the difference $E_0^+ - E_0^-$ of chiral and anti-chiral momentum modes
\cite{Ganor:1996xg,Dasgupta:1996yh,Gates:2000fj}.  

To verify this expectation, let us therefore  
compactify the 2d $(0,2)$ theory on $S^1$ and compute the vacuum energy of the matter fields along the circle, following the familiar rules:
\be
\ba
{\rm real \, periodic \,\, scalar}: E_0 &= - \frac{1}{24} \cr 
{\rm real \, periodic \,\, fermion}: E_0 &=  \frac{1}{24} 
\ea
\qquad 
\ba
 {\rm real \, anti-periodic \,\, scalar}: E_0 &=  \frac{1}{48} \cr
{\rm real \, anti-periodic \,\, fermion}: E_0 &=  - \frac{1}{48} \,.
\ea
\ee
The complex scalar in each $(0,2)$ chiral multiplet can be decomposed into 2 real chiral and anti-chiral scalars. Since we must choose the same boundary conditions for all fields inside the same $(0,2)$ multiplet, the contribution of the two chiral scalars cancels the vacuum energy of the two chiral Majorana-Weyl fermions. For periodic boundary conditions, each $(0,2)$ chiral multiplet hence contributes $-2 \times \frac{1}{24}$ to $E_0^-$. 

{Each of the $\tau (B_4)$ tensor multiplets, which contain one real anti-chiral scalar per multiplet, contributes $E_0^- = -{1\over 24}$ if we likewise assign periodic boundary conditions along $S^1$. 
The Fermi multiplets, which comprise two anti-chiral Majorana-Weyl fermions, again with periodic boundary conditions, contribute $E_0^- = + 2 \times \frac{1}{24}$.} The contribution from the moduli sector to the vacuum energy along $S^1$ is then
\be
\ba
E_0^-|_{\rm moduli} &= \frac{1}{24} \Big( - 2 N_c - \tau(B_4) + 2 (h^{2,1}(B_4) + h^{3,1}(Y_4) - h^{3,1}(B_4))  \Big) \\
& =  - \frac{2}{24} \Big( {\mathfrak a}_{\rm moduli} + \frac{3}{2} \tau(B_4)
\Big) \,,
\ea
\ee
with $N_c$ the number of $(0,2)$ chiral multiplets in table \ref{tab:SpecO}.

{In addition there are $ 8 \, c_1(B_4) \cdot C$ Fermi multiplets from the spectrum of 3-7 strings in table \ref{tab:D3spec}, each containing one complex anti-chiral fermion. The nature of these states as anti-chiral fermions follows from their origin in the open Ramond sector in perturbative setups. As all the other fields  they are assigned periodic boundary conditions along $S^1$.}
The complete D3-brane sector, including these 3-7 modes, contributes
 \be
 \ba
{ E_0^-|_{\rm D3} = \frac{1}{24} \Big( - 4 c_1(B_4) \cdot C +16 c_1(B_4) \cdot C\Big) = - \frac{2}{24} \Big( - {\mathfrak a}_{\rm D3} \Big) + c_1 (B_4) \cdot C \,.}
 \ea
 \ee
In total we therefore find
\be
\ba
{ E_0^+ - E_0^- =  - \frac{1}{24} \Big( -2 {\mathfrak a}_{\rm moduli} - 3 \tau(B_4) +  2 {\mathfrak a}_{\rm D3}  \Big)- c_1(B_4) \cdot C = \int_{B_4} \tW_4 - c_1 (B_4) \cdot C \,, }
\ea
\ee
where we have used the explicit expressions for the anomaly coefficients in terms of the characteristic classes given in section \ref{sec:GravAnF} together with $\chi_0(B_4) = 1$.
This matches precisely the number (\ref{labelnF}) of M2-branes wrapping the elliptic fiber and hence the number of winding modes after performing a T-duality.

\section{Conclusions}

In this note we determined the supergravity spectrum of 2d $(0,2)$ F-theory compactifications, using {its relation to} Type IIB string theory and {duality to} M-theory.
An important ingredient in establishing cancellation of gravitational anomalies has been the universal contribution to the anomalies from the duality twisted D3-brane sector \cite{LSNW}. Duality with M-theory is particularly subtle for two reasons. First in order to match the $S^1$ reduced particle spectra from F-theory with the massless states of the M-theory $N=2$ SQM we must invoke automorphic  duality in 1d supersymmetric theories. The latter has already been key  \cite{LSNW} in comparing the duality twisted D3-brane spectrum to that of an M2-brane, both wrapping the same curve on the F/M-theory base. 
It would be interesting to understand the 2d/1d relation better, and in particular to extend it to the full effective action of the 2d $(0,2)$ supergravity as opposed to only its massless spectrum. 
Secondly,  the M2-brane tadpole along
the elliptic fiber is responsible for a non-zero winding number of the 1d
M-theory vacuum. 
{This is reproduced from the F-theory perspective if we treat all the states with periodic boundary conditions along the $S^1$.} This complements observations on Type
IIB/M-theory duality made already in \cite{Dasgupta:1996yh}.  
{The subtleties of F/M-theory duality
hence continue to mesmerize the so-inclined.}

\subsection*{\bf Acknowledgements}

The work of CL and TW was partially supported by DFG under Grant TR33 'The
Dark Universe' and under GK 1940 'Particle Phyiscs
Beyond the Standard Model'.  SSN acknowledges support by the ERC Consolidator
Grant 682608 ``Higgs bundles: Supersymmetric Gauge Theories and Geometry
(HIGGSBNDL)".

%%%%%%%%%%%%%%%%%%%%%%%%%%%%%%%%%%%%%%%%%%%%%%%%%

%\section{Conclusions}

%%%%%%%%%%%%%%%%%%%%%%%%%%%%%%%%%%%%%%%%%%%%%%%%%
%%%%%%%%%%%%%%%%%%%%%%%%%%%%%%%%%%%%%%%%%%%%%%%%%
%%%%%%%%%%%%%%%%%%%%%%%%%%%%%%%%%%%%%%%%%%%%%%%%%
%%%%%%%%%%%%%%%%%%%%%%%%%%%%%%%%%%%%%%%%%%%%%%%%%
%%%%%%%%%%%%%%%%%%%%%%%%%%%%%%%%%%%%%%%%%%%%%%%%%
%%%%%%%%%%%%%%%%%%%%%%%%%%%%%%%%%%%%%%%%%%%%%%%%%
%\bibliography{FTheory}{}
%\bibliographystyle{JHEP} 

\def\cprime{$'$} \def\cprime{$'$}
\providecommand{\href}[2]{#2}\begingroup\raggedright\endgroup

%%%%%%%%%%%%%%%%%%%%%%%%%%%%%%%%%%%%%%%%%%%%%%%%%
%%%%%%%%%%%%%%%%%%%%%%%%%%%%%%%%%%%%%%%%%%%%%%%%%
%%%%%%%%%%%%%%%%%%%%%%%%%%%%%%%%%%%%%%%%%%%%%%%%%
%%%%%%%%%%%%%%%%%%%%%%%%%%%%%%%%%%%%%%%%%%%%%%%%%
%%%%%%%%%%%%%%%%%%%%%%%%%%%%%%%%%%%%%%%%%%%%%%%%%
%%%%%%%%%%%%%%%%%%%%%%%%%%%%%%%%%%%%%%%%%%%%%%%%%

%%%%%%%%%%%%%%%%%%%%%%%%%%%%%%%%%%%%%%%%%%%%%%%%%%
%%%%%%%%%%%%%%%%%%%%%%%%%%%%%%%%%%%%%%%%%%%%%%%%%%
%%%%%%%%%%%%%%%%%%%%%%%%%%%%%%%%%%%%%%%%%%%%%%%%%%
%\end{document}
%%%%%%%%%%%%%%%%%%%%%%%%%%%%%%%%%%%%%%%%%%%%%%%%%%
%%%%%%%%%%%%%%%%%%%%%%%%%%%%%%%%%%%%%%%%%%%%%%%%%%
%%%%%%%%%%%%%%%%%%%%%%%%%%%%%%%%%%%%%%%%%%%%%%%%%%

\end{document}